**Higher serum 25(OH)D concentration is associated with lower risk of metabolic syndrome among Aboriginal and Torres Strait Islander peoples in Australia**

Belinda Neo [1], Dale Tilbrook [2], Noel Nannup [3], John Jacky [3], Carol Michie [3], Cindy Prior [3], Eleanor Dunlop [4,5], Brad Farrant [3], Won Sun Chen [1], Carrington C.J. Shepherd [1,3,6], and Lucinda J. Black [4,5,*]

[1] Curtin Medical School, Curtin University, Bentley, Western Australia, Australia

[2] Maalingup Aboriginal Gallery, Caversham, Western Australia, Australia

[3] The Kids Research Institute Australia, The University of Western Australia, Nedlands, Western Australia, Australia

[4] Curtin School of Population Health, Curtin University, Bentley, Western Australia, Australia

[5] Institute for Physical Activity and Nutrition (IPAN), School of Exercise and Nutrition Sciences, Deakin University, Geelong, Victoria, Australia

[6] Ngangk Yira Institute, Murdoch University, Murdoch, Western Australia, Australia

*Correspondence: Lucinda J Black, Institute for Physical Activity and Nutrition (IPAN), School of Exercise and Nutrition Sciences, Deakin University, Geelong, VIC 3220, Australia. lucinda.black@deakin.edu.au. Tel.: +61 3 924 45491

**Abbreviations**

25(OH)D: 25 hydroxyvitamin-D; AATSIHS: Australian Aboriginal and Torres Strait Islander Health Survey, Core Content – risk factors and selected health conditions; ABS: Australian

Bureau of Statistics; BMI: body mass index; HDL: high density lipoprotein; NATSIHMS: National Aboriginal and Torres Strait Islander Health Measures Survey; RCTs: randomised controlled trials

**Abstract**

Although previous observational studies have shown associations between serum 25-hydroxyvitamin D (25(OH)D) concentration and metabolic syndrome, this association has not yet been investigated among Aboriginal and Torres Strait Islander peoples. We aimed to investigate the association between serum 25(OH)D concentration and metabolic syndrome and its risk factors in this population group. We used cross-sectional data from the 2012-2013 Australian Aboriginal and Torres Strait Islander Health Survey. Metabolic syndrome is defined as having ≥ 3 risk factors: elevated waist circumference, elevated triglycerides, low high-density lipoprotein (HDL) cholesterol, elevated blood pressure, or elevated fasting blood glucose. We used binomial logistic regression to test associations between serum 25(OH)D concentration and metabolic syndrome, and multiple linear regression to test associations between serum 25(OH)D concentration and each risk factor. We included the following covariates: age, sex, smoking status, education level, socio-economic status, remoteness of location, season, and body mass index (BMI). After adjusting for covariates, we found that each 10 nmol/L increase in serum 25(OH)D concentration was statistically significantly associated with a 16% lower risk of metabolic syndrome (odds ratio: 0.84, 95% confidence interval: 0.76, 0.92) and a 2.1 cm (95% confidence interval: 1.65, 2.57) lower waist circumference (BMI was not included in the model for waist circumference). We found small inverse associations between serum 25(OH)D concentration and all other risk factors except systolic blood pressure. Given that higher serum 25(OH)D concentration may confer metabolic health benefits, promoting vitamin D sufficiency may be beneficial for this population.

**Introduction**

The role of vitamin D in musculoskeletal health is well-established. Further, the wide expression of vitamin D receptors throughout the body suggests that vitamin D may contribute to physiological functions beyond musculoskeletal health[1; 2]. A review of observational studies indicated the possible association between vitamin D deficiency (serum 25-hydroxyvitamin D (25(OH)D) < 50 nmol/L) and metabolic syndrome[3]. Metabolic syndrome refers to a group of risk factors that increase the risk of developing type 2 diabetes and cardiovascular diseases[4]. However, reviews of data from randomised controlled trials (RCTs) have shown inconsistent findings relating to the causative effect of serum 25(OH)D concentration on these chronic health conditions[5; 6; 7]. These inconsistent findings can be largely attributed to limitations in the design of past RCTs, such as a lack of vitamin D deficient participants, low dose of vitamin D supplementation used, and small sample sizes[5; 6; 8].

Aboriginal and Torres Strait Islander peoples are the First Nations peoples of Australia who have inhabited Australia for over 65,000 years[9]. The colonisation of Australia around 250 years ago caused significant disruption to the traditional lifestyle and dietary habits of Aboriginal and Torres Strait Islander peoples[10]. Access to traditional foods was denied, and ration foods that were high in refined carbohydrates and saturated fats were forced upon Aboriginal and Torres Strait Islander peoples[10]. The aftermath of colonisation is still prevalent today, as evidenced by the high rates of diet-related chronic health conditions such as type 2 diabetes and cardiovascular diseases, which severely impact the life expectancy of Aboriginal and Torres Strait Islander peoples[11]. About 27% of Aboriginal and Torres Strait Islander adults aged ≥ 18 years are deficient in vitamin D[12]. The prevalence of vitamin D deficiency is higher at 39% for those living in remote areas[12]. Observational studies investigating the association between vitamin D status and chronic health conditions among Aboriginal and Torres Strait Islander peoples are limited. A notable exception is a cross-sectional study[13], which found that low serum 25(OH)D concentration was associated with diabetes risk after adjusting for cardiometabolic risk factors among Aboriginal and Torres Strait Islander peoples. However, those findings were based on a small sample of Aboriginal and Torres Strait Islander peoples (n = 592) aged ≥ 16 years recruited from only three states/territories, which may not represent the wider population. Given the potential role of vitamin D in various chronic health conditions, it is important to investigate the potential association among Aboriginal and Torres Strait Islander peoples.

Epidemiological research in Australia has often involved a deficit discourse approach when reporting health outcomes among Aboriginal and Torres Strait Islander peoples, highlighting deficiencies and implying blame on communities for health inequalities[14]. Such framing perpetuates negative connotations around poor health outcomes and the marginalisation of Aboriginal and Torres Strait Islander peoples[14; 15]. Conversely, a strength-based approach focuses on the assets of health and strengths among Aboriginal and Torres Strait Islander peoples, encouraging positive change by focusing on the benefits of a health-oriented outcome[15]. Hence, using a strength-based approach, we aimed to investigate if serum 25(OH)D concentration was associated with a reduced risk of metabolic syndrome among Aboriginal and Torres Strait Islander peoples in Australia.

## Methods

### Project governance

We obtained ethics approval from the Western Australian Aboriginal Health Ethics Committee (HREC979). Our study was conducted with the guidance of Aboriginal and

Torres Strait Islander Elders and researchers from the lands of the Whadjuk Noongar (Perth) and Yawuru Nagulagun (Broome) peoples. Aboriginal and Torres Strait Islander Elders and researchers are knowledge holders who provide important cultural context and insights. A strength-based approach was central to our study, and we applied it by highlighting the positive associative links and avoiding a deficit language. Aboriginal and Torres Strait Islander Elders and researchers were consulted throughout our manuscript development to ensure that our study reflects their perspectives.

**Study population**

We used a cross-sectional study design with data from 2012-2013 Australian Aboriginal and Torres Strait Islander Health Survey, Core Content – Risk Factors and Selected Health Conditions (AATSIHS) (n = 12,947), which is the most recent available nationally representative data for this population group. The AATSIHS was conducted between April 2012 and July 2013 and comprises the 2012-2013 National Aboriginal and Torres Strait Islander Health Survey, the 2012-2013 National Aboriginal and Torres Strait Islander Nutrition and Physical Activity Survey, and the 2012-2013 National Aboriginal and Torres Strait Islander Health Measures Survey (NATSIHMS)[16]. Of the 12,947 participants aged ≥ 2 years, 8,157 were aged ≥ 18 years, and 3,293 participated in NATSIHMS[17]. Detailed methods of the AATSIHS are available elsewhere and briefly described here[18].

The sample sizes of each survey were dependent on the aims and scope of the survey, the disaggregation and accuracy needed for survey estimates, and costs[18]. The Australian Bureau of Statistics (ABS) determined survey participants' location of residence as non-remote areas (major cities, inner regional areas, and outer regional areas) or remote areas (remote areas and very remote areas) according to the Australian Statistical Geography

Standard[18; 19]. The sample sizes for discrete Aboriginal and Torres Strait Islander communities residing in remote areas of New South Wales, Queensland, South Australia, Western Australia, and the Northern Territory were derived based on the list of communities using the 2006 census population counts and the Discrete Indigenous Communities Database[18]. The sample sizes for non-remote and remote areas of New South Wales, Queensland, South Australia, Western Australia, Northern Territory, Victoria, Tasmania, and the Australian Capital were determined using the 2011 census population counts for those available, and 2006 census population counts for the remaining[18].

The AATSIHS contains the following population data: sociodemographic, anthropometric, and biomedical. Participation in all physical and biomedical measurements was voluntary; hence, participant numbers varied for each measure.

**Sociodemographic data**

The ABS reported the age of participants as continuous data for people aged 0 - 64 years. Data were clustered into groups for those aged 65-69 years, 70-74 years, and 75+ years, due to the small number of participants in those age groups. To use age as a continuous variable, we assigned the median age of 67 years for all participants aged 65 – 69 years, and 72 years for all participants aged 70 – 74 years. For participants aged 75+ years, we used 75 years.

Participants reported their highest education level according to three categories: (i) no/primary/high school, (ii) diploma/certificates, and (iii) university. The ABS classified socio-economic status using the 2011 Index of Relative Socio-economic Disadvantage[18], which ranks regions or areas based on income, education level, unemployment, and dwellings without motor vehicles. We re-categorised the scores into quintiles. The highest quintile of 5 indicated a lesser overall disadvantage, and the lowest quintile of 1 indicated a greater overall disadvantage.

Participants aged ≥ 15 years were asked whether they were current smokers, ex-smokers, or if they had never smoked. An individual who had never smoked was defined as someone who had never smoked or smoked < 100 cigarettes or < 20 pipes, cigars, or other tobacco products. We categorised participants into two groups: (i) ex/non-smokers and (ii) current smokers.

We acknowledge the traditional calendars used by Aboriginal and Torres Strait Islander peoples. However, AATSIHS data did not allow the translation of data according to the traditional calendars relevant to individual communities. Therefore, as a proxy measure of the UV-B radiation across the year, we categorised month of blood collection using the Western calendar definitions of Australian seasons: 'spring (September–November),' 'summer (December–February),' 'autumn (March–May),' and 'winter (June–August)'[20].

**Anthropometric and blood pressure measurements**

Anthropometric measurements were collected in all participants except pregnant women[18]. Before collecting their measurements, participants were asked to remove their shoes and any heavy clothing (e.g., sweaters/coats). The height of participants was measured in centimetres using a stadiometer. The weight of participants was measured in kg using digital scales. Body mass index (BMI) was calculated as weight in kilograms divided by height in meters squared.

Waist circumference was measured in centimetres using a metal measuring tape. Low and high values for waist circumference were categorised by the ABS as 68 and 140 cm, respectively, due to small samples. To account for values lower or higher than those reported, we assigned all values of 68 as 65 cm, and all values of 140 as 150 cm. All measurements were recorded to the nearest one decimal point. A second set of anthropometric measurement was conducted on 10% of randomly selected participants. A third measurement was taken if the second measurement differed from the first by > 1 cm.

Systolic and diastolic blood pressure were taken for participants aged ≥ 5 years. All blood pressure measurements were collected using the left arm; however, in cases where measurement was not feasible (e.g., due to an injury), the right arm was used. Two blood pressure readings were taken, and the second reading was used in our analysis. If a third reading was taken, we used the average of the second and third reading. Blood pressure readings were excluded if the duplicate readings differed by > 20 mmHg.

**Biochemical measurements**

Individuals who participated in the National Aboriginal and Torres Strait Islander Health Survey and National Aboriginal and Torres Strait Islander Nutrition and Physical Activity Survey were invited to participate in the NATSIHMS. The NATSIHMS collected biomedical data under the Privacy Act 1988. Ethics approval was granted at the national level by the Australian Government Department of Health and Ageing's Departmental Ethics Committee, and at the jurisdictional levels in New South Wales, South Australia, Western Australia, Northern Territory, and Queensland Health Service Districts by their respective ethics committee(s)[18]. Detailed information regarding all biochemical measures can be found elsewhere[18].

In brief, biochemical measurements were collected from individuals aged ≥ 18 years. We used biochemical data for serum 25(OH)D concentration, triglycerides, HDL cholesterol, and fasting blood glucose plasma. Blood samples were collected at Sonic Healthcare collection clinics, or at a home visit, or at a temporary clinic at Aboriginal Medical Services, except in regional areas in South Australia and Northern Territory that were serviced by the Institute of Medical and Veterinary Science Pathology, remote areas where temporary collection sites were used, and in some areas other pathology service providers were used. The collected samples were transported to Douglass Hanly Moir Pathology in New South Wales for analysis, and each analysis was completed < 72 hours after sample collection. The highest value reported by the ABS for triglycerides was 6 mmol/L. To account for values higher than those concentrations, we assigned all values of 6 as 8 mmol/L, based on the median value between 6 and 10 mmol/L (very high triglyceride levels)[21].

Serum samples were collected and stored at -80°C. Serum 25(OH)D concentration was measured using liquid chromatography with tandem mass spectrometry assay certified to the reference measurement procedures developed by the National Institute of Standards and Technology, Ghent University, and Centres for Disease Control[18; 22]. The ABS reported serum 25(OH)D concentration measured at ≤ 15 and ≥ 130 nmol/L as 15 and 130 nmol/L, respectively, due to the small number of participants with concentrations measured at these low and high values. To account for concentrations measured at < 15 and > 130 nmol/L, we assigned all values of 15 nmol/L as 7.5 nmol/L, and all values of 130 nmol/L as 165 nmol/L, based on previously published values[12].

**Metabolic syndrome**

We used the harmonised definition for metabolic syndrome risk factors suggested by the International Diabetes Federation Task Force on Epidemiology and Prevention and international cardiovascular health representing bodies[4] and the cut-points from the ABS[18] for elevated waist circumference as follows:

(i) Elevated waist circumference ($\geq$ 94 cm in males; $\geq$ 80 cm in females)

(ii) Elevated triglycerides ($\geq$ 1.7 mmol/L)

(iii) Low high-density lipoprotein (HDL) cholesterol (< 1.0 mmol/L in males; < 1.3 mmol/L in females) or history of high cholesterol

(iv) Elevated blood pressure (systolic blood pressure $\geq$ 130 mmHg and/or diastolic blood pressure $\geq$ 85 mmHg) or history of hypertension

(v) Elevated fasting blood glucose ($\geq$ 5.6 mmol/L) or history of type 2 diabetes

The medical history of these factors was determined using the International Statistical Classification of Diseases and Related Health Problems, Tenth Revision data collected in the AATSIHS. We gave a score of 1 for the presence of each risk factor, and a final score of $\geq$ 3 was defined as metabolic syndrome.

**Statistical analysis**

All statistical analyses were conducted in the ABS DataLab, a secured environment to analyse ABS survey data[23]. Statistical analyses were performed using Stata Statistical Software: Version 18 (StataCorp, College Station, Texas) provided in the ABS DataLab. All data outputs generated in our study were approved and cleared by the ABS.

Characteristics of participants with and without metabolic syndrome were reported using descriptive statistics. Characteristic variables were visually assessed for normality. We used survey-weighted means for normally distributed continuous variables, survey-weighted medians for non-parametric continuous variables, and survey-weighted Pearson chi-square test for categorical variables, to examine group differences of participants with and without metabolic syndrome. We reported the number of participants for each metabolic syndrome risk factor, ≥ 3 metabolic syndrome risk factors, ≥ 4 metabolic syndrome risk factors, and all five metabolic syndrome risk factors. We applied a weighting variable supplied by the ABS to weight the survey data to the 2012-2013 Aboriginal and Torres Strait Islander population[18]. Due to limitations of the statistical software available in the ABS DataLab, we could not apply the weighting variable to compare between-group differences for non-parametric data.

We divided serum 25(OH)D concentration by 10 to report associations per 10 nmol/L. We used complete case analysis for survey-weighted binary logistic regression for metabolic syndrome, which means that only participants with data for the outcome measure and all specified covariates required in each model were included in the analysis. Metabolic syndrome was coded as a binary outcome: '1' for the presence of ≥ 3 metabolic syndrome risk factors and '0' for < 3 metabolic syndrome risk factors. We used complete case analysis for survey-weighted multiple linear regression for each metabolic syndrome risk factor (waist circumference, triglycerides, HDL cholesterol, systolic and diastolic blood pressure, and fasting blood glucose). Triglycerides were non-parametric and were log-transformed for statistical analysis.

We included the following covariates based on previous research: age, sex, smoking status, education level, socio-economic status, remoteness of location, season of blood collection, and BMI[12; 19; 24]. Model 1 was unadjusted; Model 2 was adjusted for age, sex, smoking status, education level, socio-economic status, remoteness of location, and season of blood collection; Model 3 was additionally adjusted for BMI (except for the outcome of waist circumference). After adjusting for covariates, we used predictive margins to plot the association between serum 25(OH)D concentration and the risk of metabolic syndrome.

All relevant diagnostics for multiple linear regression analysis (normality, multicollinearity, and linearity) were evaluated to ensure the validity of the analysis. The fit of all models was assessed using R-squared, which provides the proportion of variance in each risk factor that is explained by serum 25(OH)D concentration and the covariates. We applied a survey weight supplied by the ABS to all models[18]. Statistical significance for all models was defined as $P < 0.05$.

## Results

### Participant characteristics

Out of 3,293 participants in the biomedical component, 2,200 participants provided a fasting blood sample, of which 2,042 participants provided measurements for all 5 metabolic syndrome risk factors, and 158 participants had one or more missing risk factors measurements due to refusal or other unspecified reasons. There were 1048 participants who had ≥ 3 metabolic syndrome risk factors. Participant characteristics, stratified by metabolic syndrome, are presented in **Table 1**. Participants with metabolic syndrome were statistically significantly older, living in remote areas, and had lower serum 25(OH)D concentration, higher waist circumference, systolic blood pressure, diastolic blood pressure, triglycerides, fasting blood glucose, and lower HDL cholesterol.

Metabolic syndrome was present in about half (40.6%) of the Aboriginal and Torres Strait Islander peoples (**Table 2**). The prevalence of metabolic syndrome (≥ 3 metabolic syndrome risk factors) was similar among males (41.9%) and females (39.3%). The most common metabolic syndrome risk factor was elevated waist circumference at 72.0%, and the least prevalent risk factor was elevated fasting blood glucose or history of type 2 diabetes at 25.7%. A higher percentage of females (55.0%) had low HDL cholesterol or history of high cholesterol compared with males (34.5%). A higher percentage of males (42.0%) had elevated triglyceride levels compared with females (27.3%).

**Serum 25(OH)D concentration and metabolic syndrome**

Across all models, serum 25(OH)D concentration was statistically significantly inversely associated with metabolic syndrome **(Table 3)**. In model 3, after adjusting for all covariates, each 10 nmol/L increase in serum 25(OH)D concentration was associated with a 16% lower risk of metabolic syndrome. The association between serum 25(OH)D concentration and the predicted risk of having metabolic syndrome is depicted in **Figure 1.**

**Serum 25(OH)D concentration and waist circumference**

Across all models, serum 25(OH)D concentration was inversely associated with waist circumference. In model 2, each 10 nmol/L increase in serum 25(OH)D concentration was significantly associated with a 2.11 cm decrease in waist circumference.

**Serum 25(OH)D concentration and triglycerides**

In all models, serum 25(OH)D concentration was inversely associated with triglycerides. In model 3, after adjusting for all covariates, each 10 nmol/L increase in serum 25(OH)D concentration was significantly associated with a 3% decreased risk of elevated triglycerides.

**Serum 25(OH)D concentration and HDL cholesterol**

Across all models, there was a small statistically significant positive association between serum 25(OH)D concentration and HDL cholesterol. In model 3, after adjusting for all covariates, each 10 nmol/L increase in serum 25(OH)D concentration was significantly associated with a 0.02 mmol/L increase in HDL cholesterol.

**Serum 25(OH)D concentration and blood pressure**

In models 1 and 2, serum 25(OH)D concentration was inversely associated with systolic blood pressure. In model 3, after adjusting for all covariates, we found no associations between serum 25(OH)D concentration and systolic blood pressure.

Across all models, serum 25(OH)D concentration was inversely associated with diastolic blood pressure. In model 3, after adjusting for all covariates, each 10 nmol/L increase in serum 25(OH)D concentration was significantly associated with a 0.51 mmHg decrease in diastolic blood pressure.

**Serum 25(OH)D concentration and fasting blood glucose**

Across all models, higher serum 25(OH)D concentration was associated with a small decrease in fasting blood glucose. In model 3, after adjusting for all covariates, each 10 nmol/L increase in serum 25(OH)D concentration was significantly associated with a 0.05 mmol/L decrease in fasting blood glucose.

**Discussion**

To our knowledge, this was the first study to examine the association between serum 25(OH)D concentration and metabolic syndrome and its risk factors using nationally representative data for Aboriginal and Torres Strait Islander peoples. After adjusting for all covariates, we showed that a 10 nmol/L increase in serum 25(OH)D concentration was associated with a lower risk of having metabolic syndrome, and all its risk factors except systolic blood pressure. The most pronounced effect was observed between serum 25(OH)D concentration and metabolic syndrome, and waist circumference. There was a small effect for triglycerides, HDL cholesterol, diastolic blood pressure, and fasting blood glucose.

Importantly, after adjusting for all covariates, we showed that a 10 nmol/L increase in serum 25(OH)D concentration was associated with a clinically meaningful lower risk of metabolic syndrome. Our finding was consistent with other Australian cohort studies[19; 25], and population-representative studies from the United States and Canada[26; 27]. However, different metabolic syndrome criteria were used in those population-representative studies. Our results were similar to the Canadian study that used data from the 2007 - 2009 Canadian Health Measures survey, which showed that each 10 nmol/L increase in serum 25(OH)D concentration was associated with 14% lower risk of having metabolic syndrome[27]. That study was based on the metabolic syndrome criteria defined by the National Cholesterol Education Program Adult Treatment Panel III, which is similar to the criteria we used except that they had a higher threshold for waist circumference and fasting blood glucose[27]. A recent dose-response meta-analysis that included 23 observational studies with varying metabolic syndrome criteria also found that each 25 nmol/L increase in serum 25(OH)D concentration was associated with a 19% decreased risk of having metabolic syndrome[28].

We found that serum 25(OH)D concentration was inversely associated with waist circumference, a marker for abdominal obesity, reflecting the findings of other observational studies conducted in Australia and the United States[19; 26]. Although previous studies have confirmed the correlation between obesity and low circulating 25(OH)D concentration, the direction of the association remains unclear[29]. One of the potential pathophysiological mechanisms suggested was the volumetric dilution of 25(OH)D across serum, muscle, adipose tissue, and liver, which is higher in obese individuals, resulting in lower serum 25(OH)D concentration compared to non-obese individuals[30]. Another potential mechanism is that low circulating 25(OH)D concentration contributes to obesity: when the active form of vitamin D, 1,25 dihydroxyvitamin D (1,25(OH)$_2$D) is present, vitamin D receptors may inhibit adipogenesis[29]. Hence, low circulating 25(OH)D concentration may increase the rate of differentiation from pre-adipocytes to adipocytes.

We found an inverse association between serum 25(OH)D concentration and triglycerides and a positive association between serum 25(OH)D concentration and HDL cholesterol. High HDL cholesterol and low triglycerides profile are known to have protective benefits for cardiovascular health. However, other observational studies conducted in Australia and Canada showed an inverse association with triglycerides but no association with HDL cholesterol[19; 31]. When comparing meta-analyses of RCTs that investigated vitamin D supplementation and serum lipid profile, findings were mixed as one showed no improvement in HDL cholesterol or triglycerides[32], while the other showed improvements in triglyceride levels[33].

We found an inverse association between serum 25(OH)D concentration and diastolic blood pressure but not systolic blood pressure. Similarly, an Australian cohort study did not show any association between serum 25(OH)D concentration and systolic blood pressure[34]. Another Australian observational study that investigated the association between serum 25(OH)D concentration elevated blood pressure also did not show any association[19]. RCTs that examined the effects of vitamin D supplementation on blood pressure have yielded inconsistent results[35; 36; 37]. The Vitamin D Assessment (ViDA) trial conducted in New Zealand with a follow-up of 1.1 years showed that vitamin D supplementation lowered aortic systolic blood pressure but not brachial systolic and diastolic blood pressure in a sub-group analysis of individuals with vitamin D deficiency[35]. Comparatively, the Vitamin D3 - Omega-3 - Home exercise - Healthy aging and longevity (DO-HEALTH) trial conducted in Europe, which had a follow-up of 3 years, showed that vitamin D supplementation did not improve systolic or diastolic blood pressure[36]. A recent review of five Mendelian randomisation studies also showed no positive effects of serum 25(OH)D concentration on systolic or diastolic blood pressure[6], suggesting that vitamin D supplementation may not confer any benefits on blood pressure.

Similar to our findings, other Australian observational studies have shown inverse associations between serum 25(OH)D concentration and fasting blood glucose[19; 38]. One potential mechanism is that low serum 25(OH)D concentration may impair pancreatic β cell function and induce insulin resistance, resulting in type 2 diabetes[39; 40]. Despite the observational data and biological plausibility, a recent review of seven Mendelian randomisation studies did not find that serum 25(OH)D concentration lowered the risk of type 2 diabetes[6]. Instead, evidence from RCTs suggests that vitamin D supplementation may prevent progression from prediabetes to type 2 diabetes in non-obese individuals[41].

The strengths of our study include the use of nationally representative data collected from Aboriginal and Torres Strait Islander people in Australia. As different assays and laboratories can provide highly variable measurements of serum 25(OH)D concentration[42], a certified method using liquid chromatography with tandem mass spectrometry was used to measure serum 25(OH)D concentration. To ensure that all participants with metabolic syndrome risk factors were included in our models, we included participants who reported a medical history of type 2 diabetes, high cholesterol, and hypertension, as individuals under drug treatment to manage these chronic conditions may exhibit normal biochemical readings. We adopted a strength-based approach to guide the presentation of this data as it aims to support and empower Aboriginal and Torres Strait Islander peoples by highlighting the benefits of improving serum 25(OH)D concentration to achieve positive metabolic health outcomes. A limitation of our study is the use of cross-sectional data, as we could only report the association but not the direction of association or the causal link between serum 25(OH)D concentration and metabolic syndrome and its risk factors[43].

Our study showed that higher serum 25(OH)D concentration was associated with a lower risk of metabolic syndrome and some of its risk factors among Aboriginal and Torres Strait Islander peoples. Promoting vitamin D sufficiency, either through improving dietary intakes and/or safe sun exposure, may be beneficial for Aboriginal and Torres Strait Islander peoples.

**Acknowledgements**

**Author contributions**

LJB, CS and BN designed research; BN conducted research, analysed data and wrote the paper; ED, LJB, CS, NN, DT, JJ, WSC, BF, CP, and CM reviewed and edited the paper, LJB,

ED, CS, WSC, NN, DT and JJ provided supervision. LJB had primary responsibility for the final content. All authors have read and approved the final manuscript.

**Data Availability Statement:**

Data described in the manuscript was from the following national dataset found here: Australian Aboriginal and Torres Strait Islander Health Survey, Core Content - Risk Factors and Selected Health Conditions, 2012-13, https://www.abs.gov.au/statistics/microdata-tablebuilder/datalab

**Funding**

This study was supported by the National Health and Medical Research Council (GNT1184788). BN is supported by a Curtin Strategic Scholarship.

**Ethical Approval:** Ethics approval was granted by the Western Australian Aboriginal Health Ethics Committee (WAAHEC), reference HREC979.

**Table 1.** Characteristics of Aboriginal and Torres Strait Islander peoples stratified by presence of metabolic syndrome.[1,2,3]

|  | No metabolic syndrome (<3 risk factors present) | | Metabolic syndrome (≥ 3 risk factors present) | | P-value |
|---|---|---|---|---|---|
| n | 994 | | 1048 | | |
| Age, years (mean (SD)) | 33.96 | (11.76) | 44.14 | (16.54) | <0.001 |
| Sex (%) | | | | | 0.518 |
|    Male | 429 | (58.1) | 439 | (41.9) | |
|    Female | 565 | (60.7) | 609 | (39.3) | |
| Serum 25(OH)D concentration, nmol/L (mean (SD)) | 72.09 | (23.49) | 60.11 | (23.53) | <0.001 |
| Metabolic syndrome risk factors | | | | | |
|    Waist circumference, cm (mean (SD)) | 91.49 | (14.52) | 110.76 | (16.68) | <0.001 |
|    Triglycerides, mmol/L (median (25th,75th percentile)) | 1.1 | (0.8, 1.4) | 2 | (1.5, 2.7) | <0.001[4] |
|    HDL cholesterol, mmol/L (mean (SD)) | 1.31 | (0.26) | 1.05 | (0.26) | <0.001 |
|    Systolic blood pressure, mmHg (mean (SD)) | 114.13 | (12.33) | 130.07 | (21.33) | <0.001 |
|    Diastolic blood pressure, mmHg (mean (SD)) | 72.84 | (9.35) | 84.41 | (12.93) | <0.001 |
|    Fasting blood glucose, mmol/L (mean (SD)) | 4.81 | (0.64) | 6.25 | (2.72) | <0.001 |
| Body mass index, kg/m² (mean (SD)) | 27.04 | (5.79) | 33.75 | (7.33) | <0.001 |
| Smoking Status (%) | | | | | 0.758 |
|    Ex/non-smoker | 571 | (59.8) | 612 | (40.2) | |
|    Current | 423 | (58.5) | 436 | (41.5) | |
| Education level (%) | | | | | 0.289 |
|    No/Primary/High school | 533 | (56.3) | 642 | (43.7) | |
|    Certificate/Diploma | 384 | (62.5) | 352 | (37.5) | |
|    University | 77 | (61.4) | 54 | (38.6) | |
| Socio-economic Status (%) | | | | | 0.206 |
|    Quintile 1 | 550 | (55.8) | 635 | (44.2) | |
|    Quintile 2 | 210 | (63.3) | 202 | (36.7) | |
|    Quintile 3 | 92 | (55.0) | 98 | (45.0) | |
|    Quintile 4 | 100 | (66.9) | 87 | (33.1) | |
|    Quintile 5 | 42 | (66.2) | 26 | (33.8) | |
| Remoteness of location (%) | | | | | <0.001 |
|    Non-remote | 559 | (61.7) | 456 | (38.3) | |
|    Remote | 435 | (46.5) | 592 | (53.5) | |
| Season of blood collection (%) | | | | | 0.856 |
|    Summer | 174 | (59.1) | 181 | (40.9) | |
|    Autumn | 235 | (60.3) | 286 | (39.7) | |
|    Winter | 237 | (61.5) | 241 | (38.5) | |
|    Spring | 348 | (56.8) | 340 | (43.2) | |

25(OH)D, 25-hydroxyvitamin D; CI, confidence interval; HDL, high density lipoprotein.

[1]Data are presented as mean (standard deviation), median (25$^{th}$, 75$^{th}$ percentile) or number (%)
[2]Weighted to the 2012-2013 Aboriginal and Torres Strait Islander population
[3]To test between group differences, for continuous variables weighted survey means was used for parametric data and weighted survey medians were used for non-parametric data; for categorial variables Pearson chi-squared test was used.
[4]Between group comparison based on unweighted data due to limitations of statistical software available in ABS DataLab for non-parametric data

**Table 2.** Prevalence of metabolic syndrome and its risk factors among Aboriginal and Torres Strait Islander peoples aged ≥ 18 years.

|  | All participants | | | Males | | | Females | | |
| --- | --- | --- | --- | --- | --- | --- | --- | --- | --- |
|  | n | %[1] | 95% CI | n | %[1] | 95% CI | n | %[1] | 95% CI |
| Metabolic syndrome | | | | | | | | | |
| ≥ 3 risk factors | 1048 | 40.6 | 36.66, 44.73 | 439 | 41.9 | 35.42, 48.76 | 609 | 39.3 | 34.92, 43.85 |
| ≥ 4 risk factors | 569 | 19.0 | 16.51, 21.70 | 238 | 19.9 | 16.13, 24.35 | 331 | 18.0 | 15.01, 21.43 |
| 5 risk factors | 207 | 7.3 | 5.84, 9.11 | 85 | 7.6 | 5.43, 10.43 | 122 | 7.1 | 5.21, 9.49 |
| Elevated waist circumference | 2412 | 72.0 | 68.47, 75.30 | 844 | 65.2 | 59.33, 70.60 | 1568 | 78.8 | 74.79, 82.28 |
| Elevated triglycerides | 886 | 34.5 | 30.89, 38.39 | 434 | 42.0 | 35.66, 48.51 | 452 | 27.3 | 23.70, 31.24 |
| Low HDL cholesterol or history of high cholesterol | 1872 | 45.0 | 41.76, 48.32 | 564 | 34.5 | 29.70, 39.65 | 1308 | 55.0 | 51.12, 58.80 |
| Elevated blood pressure or history of hypertension | 1558 | 41.8 | 38.45, 45.17 | 687 | 44.9 | 39.37, 50.61 | 871 | 38.7 | 35.08, 42.45 |
| Elevated fasting blood glucose or history of type 2 diabetes | 885 | 25.7 | 22.85, 28.78 | 388 | 26.9 | 22.31, 32.14 | 497 | 24.5 | 21.23, 28.12 |

CI, confidence interval; HDL, high density lipoprotein

[1]Weighted to the 2012-2013 Aboriginal and Torres Strait Islander population

**Table 3.** Associations between serum 25(OH)D concentration, metabolic syndrome and its risk factors among Aboriginal and Torres Strait Islander peoples aged ≥ 18 years[1]

|  | Metabolic syndrome[2] | Waist circumference (cm)[3] | Triglycerides (mmol/L)[2,4] | HDL Cholesterol (mmol/L)[3] | Systolic blood pressure (mmHg)[3] | Diastolic blood pressure (mmHg)[3] | Fasting blood glucose (mmol/L)[3] |
|---|---|---|---|---|---|---|---|
| **Model 1** | | | | | | | |
| OR/ β | 0.80 | -1.95 | 0.96 | 0.03 | -0.64 | -0.93 | -0.10 |
| 95% CI | 0.74, 0.87 | -2.42, -1.49 | 0.94, 0.97 | 0.02, 0.04 | -1.22, -0.06 | -1.28, -0.57 | -0.14, -0.07 |
| P-value | <0.001 | <0.001 | <0.001 | <0.001 | 0.029 | <0.001 | <0.001 |
| R-squared | 0.045 | 0.071 | 0.041 | 0.055 | 0.008 | 0.037 | 0.021 |
| **Model 2** | | | | | | | |
| OR/ β | 0.78 | -2.11 | 0.96 | 0.03 | -0.55 | -0.87 | -0.08 |
| 95% CI | 0.70, 0.85 | -2.57, -1.65 | 0.94, 0.97 | 0.02, 0.04 | -1.09, -0.01 | -1.22, -0.52 | -0.12, -0.05 |
| P-value | <0.001 | <0.001 | <0.001 | <0.001 | 0.047 | <0.001 | <0.001 |
| R-squared | 0.153 | 0.202 | 0.164 | 0.175 | 0.198 | 0.108 | 0.146 |
| **Model 3** | | | | | | | |
| OR/ β | 0.84 | | 0.97 | 0.02 | -0.24 | -0.51 | -0.05 |
| 95% CI | 0.76, 0.92 | | 0.95, 0.99 | 0.01, 0.03 | -0.75, 0.28 | -0.83, -0.18 | -0.09, -0.02 |
| P-value | <0.001 | | 0.001 | <0.001 | 0.363 | 0.002 | 0.006 |
| R-squared | 0.273 | | 0.245 | 0.203 | 0.224 | 0.173 | 0.158 |

25(OH)D, 25-hydroxyvitamin D; β, beta coefficient; OR, Odds ratio.
[1]Weighted to the 2012-2013 Aboriginal and Torres Strait Islander population
[2]Odds-ratio
[3]Beta coefficient
[4]Log-triglyceride values were exponentiated and presented as odds ratio
Model 1: unadjusted; Model 2: age, sex, smoking status, education level, socio-economic status, remoteness of location, and season of blood collection; Model 3: model 2 + body mass index.

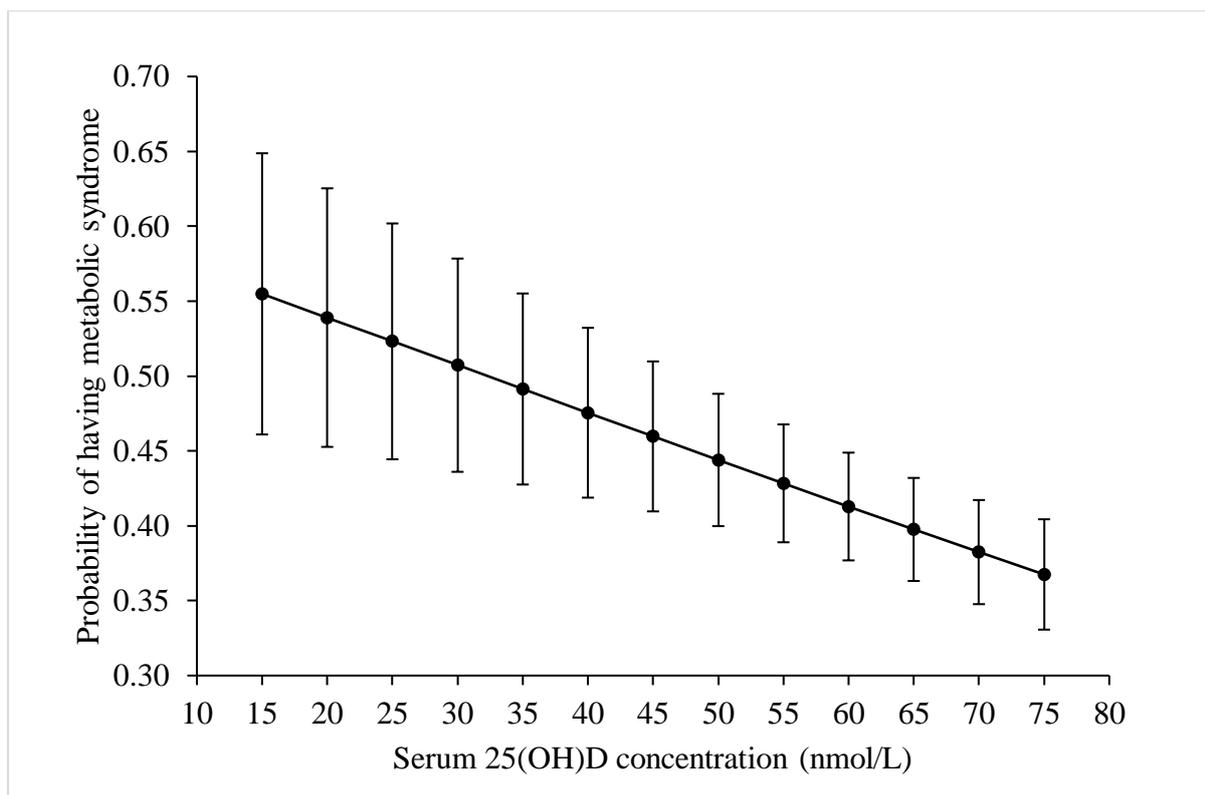

**Figure 1.** Serum 25(OH)D concentration and probability of having metabolic syndrome (defined as ≥ 3 metabolic syndrome risk factors) among Aboriginal and Torres Strait Islander peoples. The probability of having metabolic syndrome and 95% confidence interval is represented by a dot with two tails, adjusted for age, sex, smoking status, education level, socio-economic status, remoteness of location, season of blood collection, and body mass index.